\documentclass[aps,prb,floatfix,twocolumn,superscriptaddress]{revtex4-2}
\usepackage{amsmath,amssymb}
\usepackage{graphicx}
\usepackage{epsfig}
\usepackage{psfrag}
\usepackage[usenames]{color}
\usepackage{xcolor}
\usepackage{hyperref}
\usepackage{soul,color}
\usepackage{physics}
\usepackage{tabu}
\usepackage{multirow}
\usepackage{bbold}
\usepackage{mathtools}
\usepackage{siunitx}

\hypersetup{colorlinks=true,citecolor={blue},linkcolor={blue},urlcolor={blue}}

\begin{document}

\title{Theory of pseudospin resonance for multiband superconductors}

\author{Kristian~Hauser~A.~Villegas} 
\email[Corresponding author: ]{kavillegas1@up.edu.ph}
\affiliation{National Institute of Physics, University of the Philippines Diliman, Philippines.}

\begin{abstract}
We formulate a generalized pseudospin formalism for multiband superconductors in the presence of an external perturbing  electromagnetic field. Our theory naturally captures the effects of quantum band geometric quantities and is valid even for flat-band superconductors. As an interesting consequence of our theory, we show that there is an interband pairing fluctuations induced by the external field and mediated by the quantum band geometry. Surprisingly, this interband fluctuation is independent of the band gap, which can be understood from the geometric nature of such novel fluctuations. We derive the generalized equation of motion for the multiband pseudospin and the self-consistency equation. We present a formal solution to the pseudospin equation of motion in powers of the perturbing electromagnetic field. As a simple illustration of our theory, we calculate the Leggett modes for the two band case.
\end{abstract}

\maketitle

\section{Introduction}
The recent progress in twisted multilayered van der Waals materials has placed the physics of multiple flat bands into limelight. The advent of such Moire materials gave us unprecedented control over band geometry and carrier concentration \cite{Andrei2021}. By tuning the twist between one of the adjacent layers into \lq\lq magic\rq\rq angle, one can create nearly flat bands \cite{Bistritzer2011,Tarnopolsky2019}. This in turn opens the exciting possibility of studying the interplay between the band geometry and the strong electronic correlations. Here, band geometry does not merely mean band curvature but the band Berry curvature and quantum metric, which is captured by the full complex quantum geometric tensor.

In the superconducting phase of the twisted bilayer graphene \cite{Cao2018}, for example, many studies have shown that the band quantum geometry plays a central role \cite{Hu2019, Xie2020, Julku2020, Tian2023, Abouelkomsan2023}. This motivated the formulation of new theories incorporating the effects of band geometry and the existence of multibands and flat bands in a superconductor \cite{Huang2020, Shaffer2021, Chen2023}. Indeed, the presence of multiband have been shown to have nontrivial effects such as the enhancement of superfluid weight \cite{Julku2020}, breaking of time reversal symmetry \cite{Poniatowski2022}, and the existence of geometric Higgs modes \cite{Villegas2021}. 

In our previous work \cite{Villegas2021}, we have shown that the quantum band geometry leads to novel effects for the collective excitations. In particular, we have shown that the Higgs mode can be excited by an external electromagnetic field even in the flat band superconductors by virtue of band-geometric coupling. This can be relevant in the Higgs spectroscopy \cite{Chu2020} applications since this opens the possibility of using such tools for these types of superconductors. We have shown further in our previous work that there exist a second harmonic generation, in contrasts to the conventional third harmonic generation, and that the Higgs amplitude is lower bounded by the band Chern number. In this previous work, however, we did not formulate the general theory that is valid for generic multiband superconductors. In our calculations for example, we eventually focused on the single band-projected case for simplicity. The price paid for this limited scope is we missed the interesting interband effects. The aim of our current work is to fully develop a general microscopic theory of multiband  pseudospin resonance. By formulating the general multiband theory, we will show that there is a geometric interband pairing fluctuations and geometric Leggett modes that can be excited even for flat or Dirac bands. We hope that this general theory will be useful in understanding the interplay of band geometry and superconductivity in multiband systems. This will be relevant as such systems are now easily realized using Moire systems such as in twisted bilayer graphene.

This paper is organized as follows. We start our discussion with the multiorbital Hamiltonian in an external electromagnetic field in section \ref{hamiltonian}, along with the multiband Bogoliubov-de Gennes-Nambu formalism. In section \ref{expansion} we present a careful expansions of the Bloch, pairing, and interaction matrices in powers of the external field. The appearance of the quantum band geometric quantities can be traced to these expansions. In section \ref{pseudospin} we present the pseudospin formalism generalized to multiband systems. We will derive the equation of motion for the multiband pseudospin and provide the formal solution. After a careful expansion of the interaction matrix in the previous section, we show the correct form of the self-consistency equation in section \ref{selfconsistent}. We do this by first separating the particle-hole and band spaces. We then show the self-consistency equation for arbitrary choice of the Su(2N) generators, where N is the number of bands. To illustrate our multiband pseudospin theory, we provide a sample calculation in section \ref{example}, where we calculate the interband pairing and Leggett mode fluctuations. Lastly, we give the summary, conclusion, and future outlook in section \ref{summary}.
\section{Hamiltonian}
\label{hamiltonian}
\subsection{Kinetic Term}
Our starting point is the tight-binding Hamiltonian coupled to an external field via Peierls substitution
\begin{eqnarray}
\label{Hk}
H_K=\sum_{i\alpha,j\beta}\sum_\sigma\hat{c}^\dagger_{i\alpha\sigma}K^\sigma_{i\alpha,j\beta}e^{i\mathbf{A}\cdot(\mathbf{r}_{i\alpha}-\mathbf{r}_{j\beta})}\hat{c}_{j\beta\sigma},
\end{eqnarray}
where $i$ and $j$ label the lattice sites; $\alpha$ and $\beta$ label the orbitals; $\sigma$ denote spins; $K^\sigma_{i\alpha,j\beta}$ is the hopping amplitude; and $\mathbf{A}$ is the vector potential. Here we used the natural units $e=1$ and $\hbar=1$.

The Fourier transform $\Tilde{K}^\sigma(\mathbf{k})$ of $K^\sigma_{i\alpha,j\beta}$ can be diagonalized:
\begin{align}
\Tilde{K}^\sigma(\mathbf{k})=\mathcal{G}_{\mathbf{k}\sigma}\mathbb{E}_{\mathbf{k}\sigma}\mathcal{G}_{\mathbf{k}\sigma}^\dagger,
\end{align}
where $\mathbb{E}_{\mathbf{k}\sigma}\equiv diag(\varepsilon_{n\mathbf{k}\sigma})$ is a diagonal matrix composed of band dispersions $\varepsilon_{n\mathbf{k}\sigma}$ and $n$ labels the bands. The $n$-th column of the unitary matrix $\mathcal{G}_{\mathbf{k}\sigma}$ is the Bloch function of the $n$-th band. 

\subsection{Interaction and Mean-Field BCS}
We take the interaction relevant to the pairing to be of the form
\begin{align}
\label{int1}
H_{int}=&\frac{1}{2}\sum_{\mathbf{k}\mathbf{k}'}V(\mathbf{k},\mathbf{k}')_{\alpha\;\gamma}^{\;\beta\;\rho}c^{\dagger\alpha}_{\mathbf{k}}c^{\dagger}_{-\mathbf{k}\beta}c^\gamma_{-\mathbf{k}'}c_{\mathbf{k}'\rho}\\
=&\frac{1}{2}\sum_{\mathbf{k}\mathbf{k}'}V(\mathbf{k},\mathbf{k}')_{a\;c}^{\;b\;d}c^{\dagger a}_{\mathbf{k}}c^{\dagger}_{-\mathbf{k}b}c^c_{-\mathbf{k}'}c_{\mathbf{k}'d}.
\end{align}

Here the first line is written in the orbital basis, while the second line is written in the band basis. Note that we reserved the Greek letters for the orbital basis and the Roman letters for the band basis. When to raise and lower the indices will become clearer later in Subsection \ref{BdGN} when we express the mean-field BCS Hamiltonian in the Nambu formalism.\\

The mean-field pairing Hamiltonian written in the orbital space is
\begin{align}
H_{mf}=-\frac{1}{2}\sum_\mathbf{k}(\Delta_{\mathbf{k}\alpha}^{\;\;\;\beta}c^{\dagger\alpha}_\mathbf{k}c^\dagger_{-\mathbf{k}\beta}+h.c.).
\end{align}

The pairing Hamiltonian has the same form in band space with the indices $\alpha$ and $\beta$ replaced by band indices $a$ and $b$. The pairing potential must satisfy the self-consistent equation
\begin{align}
\label{sce0}
\Delta_{\mathbf{k}\alpha}^{\;\;\;\beta}=-\sum_{\mathbf{k}'}V(\mathbf{k},\mathbf{k}')_{\alpha\;\gamma}^{\;\beta\;\rho}\langle c^\gamma_{-\mathbf{k}'}c_{\mathbf{k}'\rho}\rangle.
\end{align}
The self-consistent equation in band space again has the same form and can be easily obtained by changing the orbital indices to band indices.

\subsection{Bogoliubov-de Gennes-Nambu Formalism}
\label{BdGN}
We introduce the Nambu spinor generalized for multiband systems
\begin{align}
\hat{\psi}_\mathbf{k}=(c_{1,\mathbf{k}\uparrow},\cdot\cdot\cdot,c_{N,\mathbf{k}\uparrow},c^\dagger_{1,-\mathbf{k}\downarrow},\cdot\cdot\cdot,c^\dagger_{N,-\mathbf{k}\downarrow})^T,
\end{align}
where $1,2,\cdot\cdot\cdot,N$ label the bands. The form above is written for singlet pairing. For triplet pairing, where the spins are polarized, we can simply omit the spin indices.

It will be convenient for later calculations to adopt the index notation in linear algebra and tensor analysis and and write
\begin{align}
\{c_{1,\mathbf{k}\uparrow},\cdot\cdot\cdot,c_{N,\mathbf{k}\uparrow}\}&\rightarrow c_{a,\mathbf{k}\uparrow}\nonumber\\
\{c_{1,\mathbf{k}\uparrow},\cdot\cdot\cdot,c_{N,\mathbf{k}\uparrow},c^\dagger_{1,-\mathbf{k}\downarrow},\cdot\cdot\cdot,c^\dagger_{N,-\mathbf{k}\downarrow}\}&\rightarrow c^\dagger_{a,-\mathbf{k}\downarrow}\nonumber.
\end{align}
That is, every time we write an operator with a band index such as $c_{a,\mathbf{k}\uparrow}$, we actually mean a list $c_{1,\mathbf{k}\uparrow},\cdot\cdot\cdot,c_{N,\mathbf{k}\uparrow}$. This shortens our spinor notation to
\begin{align}
\label{nambu}
\hat{\psi}_\mathbf{k}=(c_{a,\mathbf{k}\uparrow},c^\dagger_{a,-\mathbf{k}\downarrow})^T,
\end{align}
where $a$ labels the band components. 

The conjugation raises the band indices
\begin{align}
\hat{\psi}^\dagger_\mathbf{k}=(c^{\dagger a}_{\mathbf{k}\uparrow},c^a_{-\mathbf{k}\downarrow}).
\end{align}
This will proved convenient later when we write matrix multiplication as we can simply use the Einstein summation convention for the repeated covariant and contravariant indices.

The Hamiltonian can now be written in the Bogoliubov-de Gennes (BdG) form 
\begin{align}
\label{BdG1}
H=\sum_\mathbf{k}\hat{\psi}^\dagger_\mathbf{k}H_\mathbf{k}(\mathbf{A})\hat{\psi}_\mathbf{k},
\end{align}
where the Bloch Hamiltonian matrix $H_\mathbf{k}(\mathbf{A})$, upon introducing a chemical potential $\mu$, is given by
\begin{eqnarray}
\label{BdGBloch}
H_\mathbf{k}(\mathbf{A})=\frac{1}{2}
\begin{pmatrix}
\mathbb{E}_{\mathbf{k}-\mathbf{A}}-\mu & \mathcal{G}^\dagger_{\mathbf{k}-\mathbf{A}}\Delta_{\mathbf{k},orb}\mathcal{G}_{\mathbf{k}+\mathbf{A}}\\
\mathcal{G}^\dagger_{\mathbf{k}+\mathbf{A}}\Delta_{\mathbf{k},orb}^\dagger\mathcal{G}_{\mathbf{k}-\mathbf{A}} & -(\mathbb{E}_{\mathbf{k}+\mathbf{A}}-\mu)
\end{pmatrix}.
\end{eqnarray}
Here, $\Delta_{\mathbf{k},orb}$ is the pairing potential matrix in the orbital space. 

Note that eq. \eqref{BdGBloch} is written as 2$\times$2 blocks in the electron-hole space, and that the opposite charges of the electron and hole give opposite signs to the  external field $\mathbf{A}$. For example, the upper right block of eq. \eqref{BdGBloch}, in terms of explicit matrix elements, has the form
\begin{align}
\label{gaptransform}
[\Delta_{\mathbf{k},band}]_a^{\;b}=[\mathcal{G}^\dagger_{\mathbf{k}-\mathbf{A}}]_a^{\;\alpha}[\Delta_{\mathbf{k},orb}]_\alpha^{\;\beta}[\mathcal{G}_{\mathbf{k}+\mathbf{A}}]_\beta^{\; b}.
\end{align}
This is associated with the pairing $c^\dagger_{\mathbf{k}}c^\dagger_{-\mathbf{k}}$. In Nambu formalism $c^\dagger_{\mathbf{k}}$ comes from the particle degree of freedom while $c^\dagger_{-\mathbf{k}}$ comes from the hole degree of freedom. Consequently in eq. \eqref{gaptransform}, $[\Delta_{\mathbf{k},orb}]_\alpha^{\;\beta}$ transforms as a hole from the right, $[\mathcal{G}_{\mathbf{k}+\mathbf{A}}]_\beta^{\; b}$, and as a particle from the left, $[\mathcal{G}^\dagger_{\mathbf{k}-\mathbf{A}}]_a^{\;\alpha}$. From here on we will omit the label \lq\lq$orb$\rq\rq and \lq\lq$band$\rq\rq as the basis used will be clear from the type of indices (Roman or Greek) that appear.

\section{Expansion in powers of the external field}
\label{expansion}
In this section we will expand the pairing potential and the interaction matrix in powers of the external electromagnetic field. We will see that the geometric contributions appear due to the deformation of the Bloch functions. 

\subsection{Gap Function}
We begin by expanding the Bloch matrix and the band dispersion
\begin{align}
\label{expandg}
\mathcal{G}^{(\dagger)}_{\mathbf{k}\pm\mathbf{A}}&=\mathcal{G}^{(\dagger)}_\mathbf{k}\pm\partial_i\mathcal{G}^{(\dagger)}_\mathbf{k}A^i+\frac{1}{2}\partial_i\partial_j\mathcal{G}^{(\dagger)}_\mathbf{k}A^iA^j+\cdot\cdot\cdot\\
\varepsilon_{\mathbf{k}\pm\mathbf{A}}&=\varepsilon_\mathbf{k}\pm\partial_j\varepsilon_\mathbf{k}A^j+\frac{1}{2}(\partial_i\partial_j\varepsilon_\mathbf{k})A^iA^j+\cdot\cdot\cdot.
\end{align}
Substituting eq. \eqref{expandg} into the off-diagonal blocks of eq. \eqref{BdGBloch} we have, up to second order,
\begin{align}
\label{expanddelta1}
\Delta_\mathbf{k}(\mathbf{A})_a^{\;\; b}=&(\mathcal{G}^{\dagger}_\mathbf{k}-\partial_i\mathcal{G}^{\dagger}_\mathbf{k}A^i+\frac{1}{2}\partial_i\partial_j\mathcal{G}^{\dagger}_\mathbf{k}A^iA^j)_a^{\;\;\alpha}(\Delta^{(0)}_\mathbf{k}+\Delta^{(1)}_\mathbf{k}\nonumber\\
&+\Delta^{(2)}_\mathbf{k})_\alpha^{\;\;\beta}(\mathcal{G}_\mathbf{k}+\partial_i\mathcal{G}_\mathbf{k}A^i+\frac{1}{2}\partial_i\partial_j\mathcal{G}_\mathbf{k}A^iA^j)_\beta^{\;\;b}.
\end{align}

Recall that lower-case Roman letters $a$, $b$, etc. are for band indices while Greek letters are for orbital indices.

We also expand the left-hand side of the equation above
\begin{align}
\label{expanddelta2}
\Delta_\mathbf{k}(\mathbf{A})_a^{\;\; b}=\Delta^{(0)\;b}_{\mathbf{k},a}+\Delta^{(1)}_\mathbf{k}(\mathbf{A})_a^{\;\; b}+\Delta^{(2)}_\mathbf{k}(\mathbf{A})_a^{\;\; b},
\end{align}
and equate eq. \eqref{expanddelta1} and eq. \eqref{expanddelta2} order by order.

The Zeroth order is
\begin{align}
\Delta^{(0)\;b}_{\mathbf{k},a}=(\mathcal{G}_\mathbf{k}^\dagger)_a^{\;\;\alpha}(\Delta^{(0)}_\mathbf{k})_\alpha^{\;\;\beta}(\mathcal{G}_\mathbf{k})_\beta^{\;\;b},
\end{align}
which is just the usual transformation from the orbital to band basis.

The first order is given by
\begin{align}
\label{1storder1}
\Delta^{(1)}_\mathbf{k}(\mathbf{A})_a^{\;\; b}=&(\mathcal{G}_\mathbf{k}^\dagger)_a^{\;\;\alpha}(\Delta^{(1)}_{\mathbf{k}})_\alpha^{\;\;\beta}(\mathcal{G}_\mathbf{k})_\beta^{\;\;b}\nonumber\\
&-(\partial_i\mathcal{G}^\dagger_\mathbf{k})_a^{\;\;\alpha}(\Delta_\mathbf{k}^{(0)})_\alpha^{\;\;\beta}(\mathcal{G}_\mathbf{k})_\beta^{\;\;b}A^i\nonumber\\
&+(\mathcal{G}^\dagger_\mathbf{k})_a^{\;\;\alpha}(\Delta_\mathbf{k}^{(0)})_\alpha^{\;\;\beta}(\partial_i\mathcal{G}_\mathbf{k})_\beta^{\;\;b}A^i.
\end{align}

To elucidate this further, we rewrite $(\Delta^{(1)}_{\mathbf{k}})_\alpha^{\;\;\beta}$ and $(\Delta_\mathbf{k}^{(0)})_\alpha^{\;\;\beta}$ in the right hand side of eq. \eqref{1storder1} in the band basis via
\begin{align}
(\Delta^{(1)}_{\mathbf{k}})_\alpha^{\;\;\beta}=&(\mathcal{G}_\mathbf{k})_\alpha^{\;\;c}(\Delta^{(1)}_{\mathbf{k},dir})_c^{\;\;d}(\mathcal{G}^\dagger_\mathbf{k})_d^{\;\;\beta}\\
(\Delta^{(0)}_{\mathbf{k}})_\alpha^{\;\;\beta}=&(\mathcal{G}_\mathbf{k})_\alpha^{\;\;c}(\Delta^{(0)}_\mathbf{k})_c^{\;\;d}(\mathcal{G}^\dagger_\mathbf{k})_d^{\;\;\beta}.
\end{align}
Here, the subscript \lq\lq dir\rq\rq of $(\Delta^{(1)}_{\mathbf{k},dir})_c^{\;\;d}$ means the direct deformation of the order parameter due to the external field that must be calculated self-consistently. In contrasts, there will also be geometric contributions that come from the deformation of the Bloch functions as will be expounded more below in the paragraph after eq. \eqref{deltai}. We introduce the Berry connection $\mathcal{A}_{\mathbf{k}i}\equiv i\mathcal{G}^\dagger_\mathbf{k}\partial_i\mathcal{G}_\mathbf{k}$, which is in general a matrix and nonabelian for degenerate bands.

We now get
\begin{align}
\label{1storder2}
\Delta^{(1)}_\mathbf{k}(\mathbf{A})_a^{\;\; b}=\Delta^{(1)}_{\mathbf{k},dir}(\mathbf{A})_a^{\;\;b}-i\{\mathbb{\Delta}^{(0)}_\mathbf{k},\mathcal{A}_{\mathbf{k}i}\}_a^{\;\;b}A^i,
\end{align}
where the curly brackets mean anticommutator $\{A,B\}\equiv AB+BA$. Note that for general multiband systems both $\mathbb{\Delta}^{(0)}_\mathbf{k}$ and $\mathcal{A}_{\mathbf{k}i}$ are matrices.

It is convenient to factor out the explicit dependence on $A^i$ by writing
\begin{align}
\Delta^{(1)}_{\mathbf{k},dir}(\mathbf{A})_a^{\;\;b}\equiv [\Delta^{(1)}_{\mathbf{k}i,dir}]_a^{\;\;b}A^i
\end{align}
and
\begin{align}
\label{1storder3}
\Delta^{(1)}_\mathbf{k}(\mathbf{A})_a^{\;\; b}=[\Delta^{(1)}_{\mathbf{k}i}]_a^{\;\; b}A^i.
\end{align}
Equation \eqref{1storder2} now becomes
\begin{align}
\label{deltai}
[\Delta^{(1)}_{\mathbf{k}i}]_a^{\;\; b}\equiv[\Delta^{(1)}_{\mathbf{k}i,dir}]_a^{\;\;b}-i\{\mathbb{\Delta}^{(0)}_\mathbf{k},\mathcal{A}_{\mathbf{k}i}\}_a^{\;\;b}.
\end{align}

Let us now discuss the meaning and the consequence of eq. \eqref{1storder2} or eq. \eqref{deltai} above. Its second term, which comes from the second and third terms of \eqref{1storder1}, comes purely from the deformation of the Bloch functions by the external field $\partial_i\mathcal{G}^{(\dagger)}_\mathbf{k}A^i$. This deformation gives rise to the geometric contribution in first order via the Berry connection. Note that it gets multiplied only by zeroth order gap $\mathbb{\Delta}^{(0)}_\mathbf{k}$. This geometric contribution therefore does not need to be calculated self-consistently. Once we know the zeroth order pairing potential and the Berry connection, this contribution is completely determined. In contrast, first term of eq. \eqref{1storder2}, which comes from the first term of \eqref{1storder1}, does not involve derivatives of the Bloch functions. That is, the Bloch functions serve only to change from the orbital basis to the band basis. This means that this contribution comes from the \textit{direct} perturbation of the gap function due to the external field and this is the reason for the label $dir$ in \eqref{1storder2}. This term must be calculated self consistently.\\

One interesting result of our current theory comes from the second term of eq. \eqref{1storder2}. To see this, consider the case where the bands are well separated so that the zeroth-order pairing occurs only in one band, call it band $s$, where the chemical potential is located. The zeroth-order pairing can then be written as
\begin{align}
\Delta^{(0)\;b}_{\mathbf{k}a}=\delta_a^{\;\;s}\delta_s^{\;\;b}\Delta_0 f_\mathbf{k}\;\;\;\;\;\mathrm{(no\;sum\;over}\;s).
\end{align}
Now even when the initial pairing is only on band $s$, note that there are nonzero off-diagonal terms coming from the second term of \eqref{1storder2}. That is, in general, $i\{\mathbb{\Delta}^{(0)}_\mathbf{k},\mathcal{A}_{\mathbf{k}i}\}_s^{\;\;b}A^i\neq 0$ and $i\{\mathbb{\Delta}^{(0)}_\mathbf{k},\mathcal{A}_{\mathbf{k}i}\}_a^{\;\;s}A^i\neq 0$ for $a,b\neq s$. That is, there are induced interband pairings that are independent of the band gap. This is surprising as large energy difference between each electron in a Cooper pair leads to pair breaking in the conventional theory. This band gap independence can be understood as follows. The external perturbation $\mathbf{A}$ deforms the Bloch function of the $s$ band via $(\partial_i\mathcal{G}_{\mathbf{k}\alpha}^{\;\;\;s})A^i$. The resulting state becomes a superposition of different Bloch states including the other bands
\begin{align}
\partial_i\mathcal{G}_{\mathbf{k}\alpha}^{\;\;\;s}=\sum_a w_{\mathbf{k}a}\mathcal{G}_{\mathbf{k}\alpha}^{\;\;\;a},
\end{align}
where $w_{\mathbf{k}a}$ are the complex amplitudes.

Hence, the single-particle state of one of the electrons involved in the Cooper pair becomes a superposition of many Bloch states creating a band gap-independent interband pairing. This independence on the band energy scale also shows the underlying geometric nature of this induced interband pairing.

For the second order, we have the following expansion
\begin{widetext}
\begin{align}
\label{delta2nd}
\Delta^{(2)}_\mathbf{k}(\mathbf{A})_a^{\;\;b}=&(\mathcal{G}^\dagger_\mathbf{k})_a^{\;\;\alpha}(\Delta^{(2)}_\mathbf{k})_\alpha^{\;\;\beta}(\mathcal{G}_\mathbf{k})_\beta^{\;\;b}+\big[(\mathcal{G}^\dagger_\mathbf{k})_a^{\;\;\alpha}(\Delta^{(1)}_\mathbf{k})_\alpha^{\;\;\beta}(\partial_i\mathcal{G}_\mathbf{k})_\beta^{\;\;b}-(\partial_i\mathcal{G}^\dagger_\mathbf{k})_a^{\;\;\alpha}(\Delta^{(1)}_\mathbf{k})_\alpha^{\;\;\beta}(\mathcal{G}_\mathbf{k})_\beta^{\;\;b}\big]A^i\\
&+\frac{1}{2}\bigg[(\partial_i\partial_j\mathcal{G}^\dagger_\mathbf{k})_a^{\;\;\alpha}(\Delta^{(0)}_\mathbf{k})_\alpha^{\;\;\beta}(\mathcal{G}_\mathbf{k})_\beta^{\;\;b}-2(\partial_i\mathcal{G}^\dagger_\mathbf{k})_a^{\;\;\alpha}(\Delta^{(0)}_\mathbf{k})_\alpha^{\;\;\beta}(\partial_j\mathcal{G}_\mathbf{k})_\beta^{\;\;b}+(\mathcal{G}^\dagger_\mathbf{k})_a^{\;\;\alpha}(\Delta^{(0)}_\mathbf{k})_\alpha^{\;\;\beta}(\partial_i\partial_j\mathcal{G}_\mathbf{k})_\beta^{\;\;b}\bigg]A^iA^j.\nonumber
\end{align}
\end{widetext}

The first term above comes from the fluctuations of the order parameter due to the perturbation without the contributions from the deformation of the Bloch functions. Following what we did for 1st order, we can call this $\Delta^{(2)}_{\mathbf{k},dir}(\mathbf{A})_a^{\;\;b}$. The second term of \eqref{delta2nd} contains combinations of the contributions from the Bloch function deformations and first order corrections $\Delta^{(1)}_{\mathbf{k}}(\mathbf{A})$. The second line in \eqref{delta2nd} is purely geometric and comes from the second-derivative Bloch function deformations. Note that the gap function appears only in zeroth order. This means that the second line is completely determined if the zeroth order pairing and the Bloch functions are known. That is, there is no need to calculate it from a self-consistent equation.

Just like in first order, we express \eqref{delta2nd} purely in the band basis. The result is
\begin{align}
\Delta^{(2)}_\mathbf{k}&(\mathbf{A})_a^{\;\;b}=\Delta^{(2)}_{\mathbf{k},dir}(\mathbf{A})_a^{\;\;b}-i\big\{\mathbb{\Delta}^{(1)}_\mathbf{k},\mathcal{A}_{\mathbf{k}i}\big\}_a^{\;\;b}A^i\\
&-\left[\mathcal{A}_{\mathbf{k}i}\mathbb{\Delta}^{(0)}_\mathbf{k}\mathcal{A}_{\mathbf{k}j}-\frac{1}{2}(G_{\mathbf{k},ij}\mathbb{\Delta}^{(0)}_\mathbf{k}+h.c.)\right]_a^{\;\;b}A^iA^j,\nonumber
\end{align}
where 
\begin{align}
\label{G}
(G_{\mathbf{k},ij})_a^{\;\;b}\equiv(\partial_i\partial_j\mathcal{G}^\dagger_\mathbf{k})_a^{\;\;\alpha}(\mathcal{G}_\mathbf{k})_\alpha^{\;\;b}.
\end{align}

The above quantity is related to the quantum metric $g_{ij}(\mathbf{k})$ via
\begin{align}
g_{ij}(\mathbf{k})=\Tr[G_{\mathbf{k},ij}-\mathcal{A}_{\mathbf{k}i}\mathcal{A}_{\mathbf{k}j}],
\end{align}
with the trace taken in the band space.

Similar to what we did for the first order above, it is convenient to factor the external field components $A^i$ and $A^j$ explicitly. We have
\begin{align}
\label{deltasecondorder}
\Delta^{(2)}_\mathbf{k}&(\mathbf{A})_a^{\;\;b}=[\Delta_{\mathbf{k},ij}]_a^{\;\;b}A^iA^j,
\end{align}
where
\begin{align}
\label{deltaij}
[\Delta_{\mathbf{k},ij}]_a^{\;\;b}\equiv&[\Delta^{dir}_{\mathbf{k},ij}]_a^{\;\;b}-i\big\{\mathbb{\Delta}_{\mathbf{k}i},\mathcal{A}_{\mathbf{k}j}\big\}_a^{\;\;b}\\
&-\left[\mathcal{A}_{\mathbf{k}i}\mathbb{\Delta}^{(0)}_\mathbf{k}\mathcal{A}_{\mathbf{k}j}-\frac{1}{2}(G_{\mathbf{k},ij}\mathbb{\Delta}^{(0)}_\mathbf{k}+h.c.)\right]_a^{\;\;b}.\nonumber
\end{align}
The presence of two subscripts $i$ and $j$ in $[\Delta_{\mathbf{k},ij}]_a^{\;\;b}$ and $[\Delta^{dir}_{\mathbf{k},ij}]_a^{\;\;b}$ tells us that these are second order contributions and we do not have to put superscripts like the left hand side of \eqref{deltasecondorder}.

\subsection{Interaction}
\label{interaction}
Note that the geometric contributions primarily come from the transformation of the pairing potential matrix eq. \eqref{gaptransform} \cite{Peotta2015} and the expansion of the Bloch functions $\mathcal{G}_{\mathbf{k}\pm\mathbf{A}}$ in powers of $\mathbf{A}$. The complete form of this pairing potential in turn must be solved from the self-consistent equation \eqref{sce0}, which involves the matrix components of the effective electron-electron interaction $\hat{V}$. This means that we must also carefully treat the transformation and expansion of the interaction matrix $V$. 

Consider the transformation from the orbital basis to the band basis of the interaction matrix in \eqref{int1}. Recall the mean-field form and the self-consistent equation:
\begin{align}
\label{mf}
H_{mf}=&-\frac{1}{2}\sum_\mathbf{k}(\Delta_{\mathbf{k}\alpha}^{\;\;\;\beta}c^{\dagger\alpha}_\mathbf{k}c^\dagger_{-\mathbf{k}\beta}+h.c.)\\
\label{sce0b}
\Delta_{\mathbf{k}\alpha}^{\;\;\;\beta}=&-\sum_{\mathbf{k}'}V(\mathbf{k},\mathbf{k}')_{\alpha\;\gamma}^{\;\beta\;\rho}\langle c^\gamma_{-\mathbf{k}'}c_{\mathbf{k}'\rho}\rangle.
\end{align}

In going to the Nambu formalism \eqref{nambu}, we observe that the operators $c^{\dagger\alpha}_\mathbf{k}$ and $c_{\mathbf{k}'\rho}$ in eqs. \eqref{mf} and \eqref{sce0b} above, respectively, describe the electron degree of freedom. The $\alpha$ and $\rho$ indices of the interaction matrix $V(\mathbf{k},\mathbf{k}')_{\alpha\;\gamma}^{\;\beta\;\rho}$ should therefore be transformed with $[\mathcal{G}^\dagger_{\mathbf{k}-\mathbf{A}}]_a^{\;\;\alpha}$ and $[\mathcal{G}_{\mathbf{k}'-\mathbf{A}}]_\rho^{\;\;d}$, respectively. In contrasts, the operators $c^\dagger_{-\mathbf{k}\beta}$ and $c^\gamma_{-\mathbf{k}'}$ in eqs. \eqref{mf} and \eqref{sce0b} above, respectively, describe the hole degree of freedom. The $\beta$ and $\gamma$ indices of the interaction matrix $V(\mathbf{k},\mathbf{k}')_{\alpha\;\gamma}^{\;\beta\;\rho}$ should therefore be transformed using $[\mathcal{G}_{\mathbf{k}+\mathbf{A}}]_\beta^{\;\;b}$ and $[\mathcal{G}^\dagger_{\mathbf{k}'+\mathbf{A}}]_c^{\;\;\gamma}$, respectively. Note the opposite signs of $\mathbf{A}$ for the electron and the hole degrees of freedom. The momenta $\mathbf{k}$ and $\mathbf{k}'$ must also be assigned carefully.

To track clearly the electron and hole degrees of freedom, we now adopt the convention that hole indices come with bars on top: $\bar{\beta}$, $\bar{\gamma}$, $\bar{b}$, $\bar{c}$, etc.

With the observations above, we can now write the transformation
\begin{align}
V(\mathbf{k},&\mathbf{k}',\mathbf{A})_{a\;\;\bar{c}}^{\;\;\bar{b}\;\;d}\\
=&[\mathcal{G}^\dagger_{\mathbf{k}-\mathbf{A}}]_a^{\;\;\alpha}[\mathcal{G}^\dagger_{\mathbf{k}'+\mathbf{A}}]_{\bar{c}}^{\;\;\bar{\gamma}}V(\mathbf{k},\mathbf{k}')_{\alpha\;\bar{\gamma}}^{\;\bar{\beta}\;\rho}[\mathcal{G}_{\mathbf{k}+\mathbf{A}}]_{\bar\beta}^{\;\;\bar{b}}[\mathcal{G}_{\mathbf{k}'-\mathbf{A}}]_\rho^{\;\;d}.\nonumber
\end{align}
 
We now use eq. \eqref{expandg} and expand the interaction matrix in powers of the external field. 

The zeroth order is simply the transformation from the orbital basis to the band basis
\begin{align}
\label{vzero}
V^{(0)}(\mathbf{k},\mathbf{k}')_{a\;\;\bar{c}}^{\;\;\bar{b}\;\;d}&=[\mathcal{G}^\dagger_{\mathbf{k}}]_a^{\;\;\alpha}[\mathcal{G}^\dagger_{\mathbf{k}'}]_{\bar{c}}^{\;\;\bar{\gamma}}V(\mathbf{k},\mathbf{k}')_{\alpha\;\bar{\gamma}}^{\;\bar{\beta}\;\rho}[\mathcal{G}_{\mathbf{k}}]_{\bar{\beta}}^{\;\;\bar{b}}[\mathcal{G}_{\mathbf{k}'}]_\rho^{\;\;d}\nonumber\\
&\equiv V(\mathbf{k},\mathbf{k}')_{a\;\;\bar{c}}^{\;\;\bar{b}\;\;d}.
\end{align}
That is, no deformation of the Bloch functions due to the external perturbation.

The first order contribution is given by
\begin{widetext}
\begin{align}
V^{(1)}(\mathbf{k},\mathbf{k}',\mathbf{A})_{a\;\;\bar{c}}^{\;\;\bar{b}\;\;d}=-[\partial_i\mathcal{G}^\dagger_{\mathbf{k}}]_a^{\;\;\alpha}[\mathcal{G}^\dagger_{\mathbf{k}'}]_{\bar{c}}^{\;\;\bar{\gamma}}V(\mathbf{k},\mathbf{k}')_{\alpha\;\bar{\gamma}}^{\;\bar{\beta}\;\rho}[\mathcal{G}_{\mathbf{k}}]_{\bar{\beta}}^{\;\;\bar{b}}[\mathcal{G}_{\mathbf{k}'}]_\rho^{\;\;d}A^i+[\mathcal{G}^\dagger_{\mathbf{k}}]_a^{\;\;\alpha}[\partial_i'\mathcal{G}^\dagger_{\mathbf{k}'}]_{\bar{c}}^{\;\;\bar{\gamma}}V(\mathbf{k},\mathbf{k}')_{\alpha\;\bar{\gamma}}^{\;\bar{\beta}\;\rho}[\mathcal{G}_{\mathbf{k}}]_{\bar{\beta}}^{\;\;\bar{b}}[\mathcal{G}_{\mathbf{k}'}]_\rho^{\;\;d}A^i\nonumber\\
+[\mathcal{G}^\dagger_{\mathbf{k}}]_a^{\;\;\alpha}[\mathcal{G}^\dagger_{\mathbf{k}'}]_{\bar{c}}^{\;\;\bar{\gamma}}V(\mathbf{k},\mathbf{k}')_{\alpha\;\bar{\gamma}}^{\;\bar{\beta}\;\rho}[\partial_i\mathcal{G}_{\mathbf{k}}]_{\bar{\beta}}^{\;\;\bar{b}}[\mathcal{G}_{\mathbf{k}'}]_\rho^{\;\;d}A^i-[\mathcal{G}^\dagger_{\mathbf{k}}]_a^{\;\;\alpha}[\mathcal{G}^\dagger_{\mathbf{k}'}]_{\bar{c}}^{\;\;\bar{\gamma}}V(\mathbf{k},\mathbf{k}')_{\alpha\;\bar{\gamma}}^{\;\bar{\beta}\;\rho}[\mathcal{G}_{\mathbf{k}}]_{\bar{\beta}}^{\;\;\bar{b}}[\partial_i'\mathcal{G}_{\mathbf{k}'}]_\rho^{\;\;d}A^i.
\end{align}
\end{widetext}

It is convenient to express all the factors above in the band basis. In doing so, we will see the explicit appearance of the quantum band geometric quantities. We do this by inverting \eqref{vzero}. That is, expressing $V(\mathbf{k},\mathbf{k}')_{\alpha\;\bar{\gamma}}^{\;\bar{\beta}\;\rho}$ in terms of $V(\mathbf{k},\mathbf{k}')_{a\;\;\bar{c}}^{\;\;\bar{b}\;\;d}$, then using the relations
\begin{align}
\mathcal{A}^{\;\;\;\;\bar{b}}_{\mathbf{k}ja}=&i[\mathcal{G}^{\dagger}_\mathbf{k}]_a^{\;\;\alpha}[\partial_j\mathcal{G}_\mathbf{k}]_\alpha^{\;\;\bar{b}}\\
[\partial_j\mathcal{G}^{\dagger}_\mathbf{k}]_a^{\;\;\alpha}[\mathcal{G}_\mathbf{k}]_\alpha^{\;\;\bar{b}}=&-[\mathcal{G}^{\dagger}_\mathbf{k}]_a^{\;\;\alpha}[\partial_j\mathcal{G}_\mathbf{k}]_\alpha^{\;\;\bar{b}}.
\end{align}

The result is
\begin{widetext}
\begin{align}
\label{v1}
V^{(1)}(\mathbf{k},\mathbf{k}')_{a\;\;\bar{c}}^{\;\;\bar{b}\;\;d}=&-i[\mathcal{A}_{\mathbf{k}j}]_a^{\;\;e}V(\mathbf{k},\mathbf{k}')_{e\;\;\bar{c}}^{\;\;\bar{b}\;\;d}A^j+i[\mathcal{A}_{\mathbf{k}'j}]_{\bar{c}}^{\;\;\bar{f}}V(\mathbf{k},\mathbf{k}')_{a\;\;\bar{f}}^{\;\;\bar{b}\;\;d}A^j
-iV(\mathbf{k},\mathbf{k}')_{a\;\;\bar{c}}^{\;\;\bar{g}\;\;d}[\mathcal{A}_{\mathbf{k}j}]_{\bar{g}}^{\;\;\bar{b}}A^j\nonumber\\
&+iV(\mathbf{k},\mathbf{k}')_{a\;\;\bar{c}}^{\;\;\bar{b}\;\;h}[\mathcal{A}_{\mathbf{k}j}]_h^{\;\;d}A^j.
\end{align}
\end{widetext}

Here we see that the external field changes the interaction matrix via the Berry connection. 

Similar manipulations can be made to get the second order expansion of the interaction matrix. We have
\begin{widetext}
\begin{align}
\label{v2}
V^{(2)}(\mathbf{k},\mathbf{k}')_{a\;\;\bar{c}}^{\;\;\bar{b}\;\;d}=&\bigg\{V(\mathbf{k},\mathbf{k}')_{a\;\;\bar{c}}^{\;\;\bar{r}\;\;s}[\mathcal{A}_{\mathbf{k}j}]_{\bar{r}}^{\;\;\bar{b}}[\mathcal{A}_{\mathbf{k}'i}]_s^{\;\;d}-V(\mathbf{k},\mathbf{k}')_{m\;\;\bar{c}}^{\;\;\bar{r}\;\;d}[\mathcal{A}_{\mathbf{k}j}]_a^{\;\;m}[\mathcal{A}_{\mathbf{k}'i}]_{\bar{r}}^{\;\;\bar{b}}+V(\mathbf{k},\mathbf{k}')_{m\;\;\bar{c}}^{\;\;\bar{b}\;\;s}[\mathcal{A}_{\mathbf{k}j}]_a^{\;\;m}[\mathcal{A}_{\mathbf{k}'i}]_s^{\;\;d}\nonumber\\
&+V(\mathbf{k},\mathbf{k}')_{a\;\;\bar{n}}^{\;\;\bar{r}\;\;d}[\mathcal{A}_{\mathbf{k}j}]_{\bar{r}}^{\;\;\bar{b}}[\mathcal{A}_{\mathbf{k}'i}]_{\bar{c}}^{\;\;\bar{n}}-V(\mathbf{k},\mathbf{k}')_{a\;\;\bar{n}}^{\;\;\bar{b}\;\;s}[\mathcal{A}_{\mathbf{k}'i}]_{\bar{c}}^{\;\;\bar{n}}[\mathcal{A}_{\mathbf{k}'j}]_s^{\;\;d}+V(\mathbf{k},\mathbf{k}')_{m\;\;\bar{n}}^{\;\;\bar{b}\;\;d}[\mathcal{A}_{\mathbf{k}i}]_a^{\;\;m}[\mathcal{A}_{\mathbf{k}'j}]_{\bar{c}}^{\;\;\bar{n}}\nonumber\\
&+\frac{1}{2}(G^\dagger_{\mathbf{k},ij})_{\bar{r}}^{\;\bar{b}}
V(\mathbf{k},\mathbf{k}')_{a\;\;\bar{c}}^{\;\;\bar{r}\;\;d}+\frac{1}{2}
(G^\dagger_{\mathbf{k}',ij})_{s}^{\;d}V(\mathbf{k},\mathbf{k}')_{a\;\;\bar{c}}^{\;\;\bar{b}\;\;s}+\frac{1}{2}(G_{\mathbf{k},ij})_{a}^{\;m}V(\mathbf{k},\mathbf{k}')_{m\;\;\bar{c}}^{\;\;\bar{b}\;\;d}\nonumber\\
&+\frac{1}{2}(G_{\mathbf{k}',ij})_{\bar{c}}^{\;\bar{n}}
V(\mathbf{k},\mathbf{k}')_{a\;\;\bar{n}}^{\;\;\bar{b}\;\;d}
\bigg\}A^iA^j
\end{align}
\end{widetext}
where $G_{\mathbf{k},ij}$ is defined in \eqref{G}.

Equations \eqref{vzero}, \eqref{v1}, and \eqref{v2} will be useful when we solve the self-consistent equation perturbatively in powers of the external eletromagnetic vector potential.

\section{The Pseudospin Formalism}
\label{pseudospin}
After laying down the Hamiltonian and the expansions of the pairing and interaction matrices, we now proceed to the main section of this work. Here, we will generalize Anderson's pseudospin formalism \cite{Anderson1958} to include the multiband systems. We will rewrite the BdG Hamiltonian in terms of the multiband pseudospins formalism. We provide an explicit formula for the pseudomagnetic field, then derive the equations of motion for the pseudospins. We then provide a formal solution to the equations of motion using Laplace transformation and perturbation.

\subsection{Equations of motion}
We start with the generalized definition for the pseudospin of a multiband system:
\begin{align}
\vec{\Lambda}_\mathbf{k}=\hat{\psi}^\dagger_\mathbf{k}\vec{\mathbb{\Gamma}}\hat{\psi}_\mathbf{k}.
\end{align}
We note that at this stage what we have is a pseudospin \textit{operator}. This will become a $c$ number later on when we take the expectation value after we derived the equations of motion.

Here the vector components of $\vec{\mathbb{\Gamma}}$ are the generators of SU(2N), $\{\mathbb{\Gamma}^m\}$ for $m=1,\cdot\cdot\cdot, 4N^2-1$, where $N$ is the number of bands. The factor of two in SU(2N) comes from the electron and hole degrees of freedom in the Nambu formalism. Sometimes it is convenient to include the $2N\times 2N$ identity matrix. In such a case we will designate it as $\mathbb{\Gamma}^0\equiv \mathbb{I}_{2N\times 2N}$. 

Furthermore, it is standard to choose the generators that satisfy the relations
\begin{align}
\label{generatorrelations}
\mathrm{Tr} [\mathbb{\Gamma}^m\mathbb{\Gamma}^n]=\frac{1}{2}\delta^{mn},\;\;\;\;\;\;  [\mathbb{\Gamma}^l,\mathbb{\Gamma}^n]=if^{lmn}\mathbb{\Gamma}^n,
\end{align}
where $f^{lmn}$ are the structure coefficients of su($2N$) algebra. We will assume that this is the case in this work.

In terms of the pseudospins, the Hamiltonian can be written as
\begin{align}
\label{BdGpseudo}
H(\mathbf{A})=2\sum_\mathbf{k}\vec{B}_\mathbf{k}(\mathbf{A})\cdot\vec{\Lambda}_\mathbf{k},
\end{align}
where $\vec{B}_\mathbf{k}(\mathbf{A})$ is the pseudomagnetic field, whose explicit components will be solved below. For an $N$-band superconductor, this pseudomagnetic field is a $4N^2-1$-dimensional vector.

Comparing the BdG Hamiltonian eq. \eqref{BdG1} with the Hamiltonian eq. \eqref{BdGpseudo} above gives us the Bloch BdG Hamiltonian in terms of the pseudomagnetic field:
\begin{align}
\label{bdgandpseudofield}
H_\mathbf{k}(\mathbf{A})=2B_\mathbf{k}(\mathbf{A})_n \mathbb{\Gamma}^n.
\end{align}

We can express the pseudomagnetic field in terms of this BdG Hamiltonian matrix $H_\mathbf{k}(\mathbf{A})$ by multiplying eq. \eqref{bdgandpseudofield} by $\mathbb{\Gamma}^m$, taking the trace, then using the relations \eqref{generatorrelations} to get
\begin{align}
\label{pseudomagneticfield}
B_\mathbf{k}(\mathbf{A})^n=\mathrm{Tr}[\mathbb{\Gamma}^nH_\mathbf{k}(\mathbf{A})].
\end{align}
Hence, once the form of $H_\mathbf{k}(\mathbf{A})$ is known, we can calculate the components of the pseudomagnetic field vector.

The pseudospin fluctuations due to the external electromagnetic field is usually solved using perturbation. To prepare for this, let us expand the pseudomagnetic field in powers of $\mathbf{A}$. Using equations \eqref{expandg}, \eqref{1storder3}, and \eqref{deltasecondorder} we can write the BdG Hamiltonian as
\begin{align}
H_\mathbf{k}(\mathbf{A})=H^{(0)}(\mathbf{k})+H_i(\mathbf{k})A^i+H_{ij}(\mathbf{k})A^iA^j,
\end{align}
where
\begin{eqnarray}
H^{(0)}(\mathbf{k})=&\frac{1}{2}
\begin{pmatrix}
\mathbb{E}_\mathbf{k}-\mu & \mathbb{\Delta}^{(0)}_\mathbf{k}\\
\mathbb{\Delta}^{(0)\dagger}_\mathbf{k} & -\mathbb{E}_\mathbf{k}+\mu
\end{pmatrix},\\
H_i(\mathbf{k})=&\frac{1}{2}
\begin{pmatrix}
\;\;\partial_i\mathbb{E}_\mathbf{k}\;\; & \mathbb{\Delta}_{\mathbf{k}i}\\
\mathbb{\Delta}^{\dagger}_{\mathbf{k}i} & \;\;-\partial_i\mathbb{E}_\mathbf{k}\;\;
\end{pmatrix},\\
H_{ij}(\mathbf{k})=&\frac{1}{2}
\begin{pmatrix}
\frac{1}{2}\partial_i\partial_j\mathbb{E}_\mathbf{k}\;\; & \mathbb{\Delta}_{\mathbf{k}ij}\\
\mathbb{\Delta}^{\dagger}_{\mathbf{k}ij} & \;\;-\frac{1}{2}\partial_i\partial_j\mathbb{E}_\mathbf{k}\;\;
\end{pmatrix}.
\end{eqnarray}

Here, $\mathbb{\Delta}_{\mathbf{k}i}$ and $\mathbb{\Delta}_{\mathbf{k}ij}$ are defined in eqs. \eqref{deltai} and \eqref{deltaij}, respectively.

The nth component of the pseudomagnetic field up to second order can then be written as
\begin{align}
\label{bcomponent0}
B^{(0)n}_\mathbf{k}=&\Tr[\mathbb{\Gamma}^nH^{(0)}(\mathbf{k})]\\
\label{bcomponent1}
B^{(1)n}_\mathbf{k}=&\Tr[\mathbb{\Gamma}^nH_i(\mathbf{k})]A^i\equiv B^n_i(\mathbf{k})A^i\\
\label{bcomponent2}
B^{(2)n}_\mathbf{k}=&\Tr[\mathbb{\Gamma}^nH_{ij}(\mathbf{k})]A^iA^j\equiv B^n_{ij}(\mathbf{k})A^iA^j.
\end{align}

The equations of motion can be calculated from the Heisenberg equation
\begin{align}
\label{eom1}
\partial_t\Lambda^l_\mathbf{k}=i[H(\mathbf{A}),\Lambda^l_\mathbf{k}].
\end{align}
Note that even though we have rewritten the Hamiltonian in terms of the pseudospin, in order for the commutator at the right hand side to make sense, the pseudospin must be an operator at this stage. An alternative is to replace the commutation relation by the Poisson bracket. The pseudospin can then be replaced immediately by its semicalssical value.

To evaluate the commutator at the right hand side of eq. \eqref{eom1} we first note that the multiband Nambu spinor \eqref{nambu} obeys the anticommutation relations
\begin{align}
\label{anticommutation}
\big\{\hat{\psi}_\mathbf{k},\hat{\psi}^\dagger_{\mathbf{k}'}\big\}=\delta_{\mathbf{kk}'}\mathbb{I}\;\;\mbox{and}\;\;\big\{\hat{\psi}_\mathbf{k},\hat{\psi}_{\mathbf{k}'}\big\}=0,
\end{align}
where $\mathbb{I}$ is now a $2N\times 2N$ identity matrix.

Using the Hamiltonian \eqref{BdGpseudo} the Heisenberg equation \eqref{eom1} becomes
\begin{align}
\partial_t\Lambda^l_\mathbf{k}&=2i\sum_{\mathbf{k}'}\big[B^j_{\mathbf{k}'}(\mathbf{A})\Lambda_{\mathbf{k}'j},\Lambda_{\mathbf{k}i}\big]\\
\label{eom2}
&=2i\sum_{\mathbf{k}'}B^j_{\mathbf{k}'}(\mathbf{A})\big[\Lambda_{\mathbf{k}'j},\Lambda_{\mathbf{k}i}\big].
\end{align}

We therefore need to evaluate the commutator between the two pseudospin operators
\begin{align}
\big[\Lambda_{\mathbf{k}'j},\Lambda_{\mathbf{k}i}\big]
=\big[\hat{\psi}^\dagger_{\mathbf{k}'}\Gamma_j\hat{\psi}_{\mathbf{k}'},\hat{\psi}^\dagger_{\mathbf{k}}\Gamma_i\hat{\psi}_{\mathbf{k}}\big].
\end{align}

It is convenient to rewrite this in terms of the SU(2N) indices, which we will denote by $A$, $B$, $C$, and $D$, and use the Einstein summation convention
\begin{align}
\big[\Lambda_{\mathbf{k}'j},\Lambda_{\mathbf{k}i}\big]
=&\big[\hat{\psi}^{\dagger\;A}_{\mathbf{k}'}\Gamma_{jA}^B\hat{\psi}_{\mathbf{k}'B},\hat{\psi}^{\dagger\;C}_{\mathbf{k}}\Gamma_{iC}^D\hat{\psi}_{\mathbf{k}D}\big]\\
=&\Gamma_{jA}^B\Gamma_{iC}^D\big[\hat{\psi}^{\dagger\;A}_{\mathbf{k}'}\hat{\psi}_{\mathbf{k}'B},\hat{\psi}^{\dagger\;C}_{\mathbf{k}}\hat{\psi}_{\mathbf{k}D}\big]\\
\label{parenthesis}
=&\Gamma_{jA}^B\Gamma_{iC}^D\big(\hat{\psi}^{\dagger\;A}_{\mathbf{k}'}\hat{\psi}_{\mathbf{k}'B}\hat{\psi}^{\dagger\;C}_{\mathbf{k}}\hat{\psi}_{\mathbf{k}D}\nonumber\\
&-\hat{\psi}^{\dagger\;C}_{\mathbf{k}}\hat{\psi}_{\mathbf{k}D}\hat{\psi}^{\dagger\;A}_{\mathbf{k}'}\hat{\psi}_{\mathbf{k}'B}\big)
\end{align}

Now consider the second term inside the parenthesis of the last line in eq. \eqref{parenthesis}. We bring the factor $\hat{\psi}^{\dagger\;A}_{\mathbf{k}'}\hat{\psi}_{\mathbf{k}'B}$ to the left of $\hat{\psi}^{\dagger\;C}_{\mathbf{k}}\hat{\psi}_{\mathbf{k}D}$ by using the anitcommutation relations for the multiband Nambu spinor eq. \eqref{anticommutation}. The result is
\begin{align}
\big[\Lambda_{\mathbf{k}'j},\Lambda_{\mathbf{k}i}\big]
=&-\delta_{\mathbf{kk}'}\hat{\psi}^{\dagger\;C}_\mathbf{k}\Lambda_{iC}^A\Lambda_{jA}^B\hat{\psi}_{\mathbf{k}'B}\nonumber\\
&+\delta_{\mathbf{kk}'}\hat{\psi}^{\dagger\;A}_{\mathbf{k}'}\Gamma_{jA}^B\Gamma_{iB}^D\hat{\psi}_{\mathbf{k}D}.
\end{align}

For the first term, we change the dummy indices $C\rightarrow A$, $A\rightarrow B$, and $B\rightarrow D$. Furthermore, because of the Dirac delta $\delta_{\mathbf{kk}'}$ we can replace the momentum subscripts of the Nambu spinors $\mathbf{k}'\rightarrow \mathbf{k}$. The equation above then becomes
\begin{align}
\big[\Lambda_{\mathbf{k}'j},\Lambda_{\mathbf{k}i}\big]
=&\delta_{\mathbf{kk}'}\hat{\psi}^{\dagger\;A}_\mathbf{k}(\Gamma_{jA}^B\Gamma_{iB}^D-\Gamma_{iA}^B\Gamma_{jB}^D)\hat{\psi}_{\mathbf{k}D}.
\end{align}

Note that the indices has the correct placing, i.e. paired up and down and adjacent, so that we can rewrite the right hand side above as matrix multiplications. The difference then gives the commutator
\begin{align}
\label{lambdacommutator}
\big[\Lambda_{\mathbf{k}'j},\Lambda_{\mathbf{k}i}\big]
=&\delta_{\mathbf{kk}'}\hat{\psi}^{\dagger}_{\mathbf{k}}\big[\Gamma_j,\Gamma_i\big]\hat{\psi}_\mathbf{k}.
\end{align}

Now recall that $\Gamma_i$s are the generators of SU(2N) and obeys the Lie algebra
\begin{align}
\big[\Gamma_j,\Gamma_i\big]=if_{jin}\Gamma_n.
\end{align}

Using this into eq.\eqref{lambdacommutator} we have
\begin{align}
\big[\Lambda_{\mathbf{k}'j},\Lambda_{\mathbf{k}i}\big]
=&i\delta_{\mathbf{kk}'}f_{jin}\hat{\psi}^{\dagger}_{\mathbf{k}}\Gamma_n\hat{\psi}_\mathbf{k}=i\delta_{\mathbf{kk}'}f_{jin}\Lambda_{\mathbf{k}n}.
\end{align}

Finally, substituting the result above into eq.\eqref{eom2} we obtain the equations of motion
\begin{align}
\partial_t\Lambda_{\mathbf{k}l}=2f_{ijn}B^j_\mathbf{k}(\mathbf{A})\Lambda_{\mathbf{k}n}.
\end{align}
We can now replace the pseudospin operators by the classical values or take the expectation values as is done in the original pseudospin formalism \cite{Anderson1958}.

\subsection{Formal solution}
Having derived the equations of motion for the general multiband pseudospin, we now provide its formal solution using perturbation and Laplace transformation. Note that the order matters: one must first expand in powers of $\mathbf{A}$, along with 
\begin{align}
\label{pseudomagneticfieldexpand}
\mathbf{B}_\mathbf{k}(\mathbf{A})=\mathbf{B}_\mathbf{k}^{(0)}+\mathbf{B}_\mathbf{k}^{(1)}+\mathbf{B}_\mathbf{k}^{(2)}+\cdot\cdot\cdot,
\end{align}
where the components are given by the equations \eqref{bcomponent0}, \eqref{bcomponent1}, and \eqref{bcomponent2},
and
\begin{align}
\mathbf{\Lambda}_\mathbf{k}=\Lambda_\mathbf{k}^{(0)}+\Lambda_\mathbf{k}^{(1)}+\Lambda_\mathbf{k}^{(2)}+\cdot\cdot\cdot,
\end{align}
where the superscripts reflect the orders in powers of $\mathbf{A}$.

The Laplace transformation must then be used to solve the differential equation order by order.

 For zeroth order we have
\begin{align}
\label{zerothorder}
0=2f_{mn}^{\;\;\;\;\;l}B^{(0)}_\mathbf{k}(\mathbf{A})^{m}\Lambda_\mathbf{k}^{(0)n}.
\end{align}

Since the structure coefficients are antisymmetric under the exchange of the $m$ and $n$ indices $f_{mn}^{\;\;\;\;\;l}=-f_{nm}^{\;\;\;\;\;l}$, this shows that $\Lambda_\mathbf{k}^{(0)n}\propto B^{(0)}_\mathbf{k}(\mathbf{A})^{n}$ is a solution. Since we want the energy to be minimized at the ground state, we choose the solution to be antiparallel to the pseudomagnetic field
\begin{align}
\label{lamda0}
\vec{\Lambda}_\mathbf{k}^{(0)}=-\hat{B}^{(0)}_\mathbf{k}(\mathbf{A}).
\end{align}

This solution is reasonable as our Hamiltonian \eqref{BdGpseudo} is analogous to a system of spin immersed in a magnetic field but in higher dimensions.

Let us now deal with the higher order equations of motion. Unlike the zeroth order eq. \eqref{zerothorder}, these higher orders contain derivatives with respect to time and are therefore proper differential equations. We assume that the perturbing external field vanishes in the far past $\mathbf{A}(t\rightarrow -\infty) =0$, so that we can take the initial condition to be $\Lambda^{(i)}(t\rightarrow -\infty)=0$ for $i\geq 1$. 

The Laplace transform of the first order equation of motion is
\begin{align}
-s\Lambda_\mathbf{k}^{(1)l}=2f_{mn}^{\;\;\;\;\;l}\left(B^{(0)}_\mathbf{k}(\mathbf{A})^{m}\Lambda_\mathbf{k}^{(1)n}+B^{(1)}_\mathbf{k}(\mathbf{A})^{m}\Lambda_\mathbf{k}^{(0)n}\right),
\end{align}
where $s$ comes from the Laplace transform, which trades off $t\rightarrow s$.

This is just a system of algebraic equations (for $l=1,2,\cdot\cdot\cdot 4N^2-1$), which we can solve for each components $\Lambda_\mathbf{k}^{(1)j}$ giving
\begin{align}
\label{pseudospinsolution1}
\Lambda_\mathbf{k}^{(1)j}=-2(M_0^{-1})^j_{\;l}f^l_{\;mn}B^{(1)}_\mathbf{k}(\mathbf{A})^{m}\Lambda_\mathbf{k}^{(0)n},
\end{align}
where $M_0^{-1}$ is the inverse of the matrix $M_0$ whose elements are given by
\begin{align}
(M_0)^l_{\;n}\equiv s\delta^l_{\;n}+2f^l_{\;mn}B^{(0)}_\mathbf{k}(\mathbf{A})^m.
\end{align}
Note that by the time that we are calculating the first order solution \eqref{pseudospinsolution1}, the zeroth order solution is already known from eq. \eqref{lamda0}.

Once the first order correction to the pseudospin is known, the same steps can be done for the second order. This gives
\begin{align}
\label{pseudospinsolution2}
\Lambda_\mathbf{k}^{(2)j}=-2(M_0^{-1})^j_{\;l}f^l_{\;mn}\left(B^{(1)}_\mathbf{k}(\mathbf{A})^{m}\Lambda_\mathbf{k}^{(1)n}+B^{(2)}_\mathbf{k}(\mathbf{A})^{m}\Lambda_\mathbf{k}^{(0)n}\right).
\end{align}

Equations \eqref{lamda0}, \eqref{pseudospinsolution1}, and \eqref{pseudospinsolution2} give us the formal solutions to the pseudospin fluctuations up to second order in the external field.

\section{The Self-consistent Equation}
\label{selfconsistent}
In the previous section, we derived equations \eqref{lamda0}, \eqref{pseudospinsolution1}, and \eqref{pseudospinsolution2} and called them \lq\lq formal solutions\rq\rq. There is, however, a caveat: this assumes that the pseudomagnetic field components are known. In actuality, the pseudomagnetic field depends on the pairing potential, which in turn, is to be calculated using the expectation value eq. \eqref{sce0}. The complete system of equations to be solved therefore includes a self-consistent equation in addition to the equations of motion for the pseudospins.

In this section, we will show how to correctly write the self-consistent equation in terms of the generalized multiband pseudospin formalism. This is where our expansion of the interaction potential matrix that we did in section \ref{interaction} becomes useful.

We start with the self-consistent equation analogous to eq. \eqref{sce0} but written in the band basis
\begin{align}
\label{sce1}
\Delta_\mathbf{k}(\mathbf{A})_a^{\;\;\bar{b}}=&-\sum_{\mathbf{k}'}V(\mathbf{k},\mathbf{k}',\mathbf{A})_{a\;\;\bar{c}}^{\;\;\bar{b}\;\;d}\langle c^{\bar{c}}_{-\mathbf{k}'}c_{\mathbf{k}'d}\rangle
\end{align}
Recall from our discussion in section \ref{interaction} that a bar on top of an index denotes a hole degree of freedom.

We want to express the self-consistency condition \eqref{sce1} in terms of the pseudospin. To do this we write the interaction as a matrix
\begin{align}
\label{vmatrix}
\left[V(\mathbf{k},\mathbf{k}',\mathbf{A})_a^{\;\bar{b}}\right]_{\bar{c}}^{\;d},
\end{align}
where $\bar{c}$ and $d$ label the elements of the matrix $\mathbb{V}(\mathbf{k},\mathbf{k}',\mathbf{A})_a^{\;\bar{b}}$ for fixed $a$ and $\bar{b}$. For example, $\left[V(\mathbf{k},\mathbf{k}',\mathbf{A})_1^{\;\bar{2}}\right]_{\bar{3}}^{\;4}$ means the third row and fourth column of the particular matrix labeled by $a=1$ and $\bar{b}=\bar{2}$. 

We expand these matrices in terms of the generators of SU(2N) (including the identity matrix) and write
\begin{align}
\label{vlambdaexpand}
\left[V(\mathbf{k},\mathbf{k}',\mathbf{A})_a^{\;\bar{b}}\right]_{\bar{c}}^{\;d}
=\vec{V}(\mathbf{k},\mathbf{k}',\mathbf{A})_a^{\;\bar{b}}\cdot\left[\vec{\mathbb{\Gamma}}\right]_{\bar{c}}^{\;d},
\end{align}
where
\begin{align}
\label{vvector}
\vec{V}(\mathbf{k},\mathbf{k}',\mathbf{A})_a^{\;\bar{b}}\equiv\Tr\left\{\vec{\mathbb{\Gamma}}\mathbb{V}(\mathbf{k},\mathbf{k}',\mathbf{A})_a^{\;\bar{b}}\right\}.
\end{align}

Note that the above equations are just the generalization of the well known expansion of any two by two matrix in terms of the Pauli matrices, but with the Pauli matrices replaced by the SU(2N) group generators. The components of $\vec{V}(\mathbf{k},\mathbf{k}',\mathbf{A})_a^{\;\bar{b}}$ are the just the number coefficients. We also clarify the possible confusion of notation here: $\vec{V}(\mathbf{k},\mathbf{k}',\mathbf{A})_a^{\;\bar{b}}$ is different from $\mathbb{V}(\mathbf{k},\mathbf{k}',\mathbf{A})_a^{\;\bar{b}}$. The former is defined via eq. \eqref{vmatrix} and has a matrix structure, while the latter is defined via eq. \eqref{vvector} and has a vector structure.

Using eq. \eqref{vlambdaexpand} into eq. \eqref{sce1} we get
\begin{align}
\Delta_\mathbf{k}(\mathbf{A})_a^{\;\;\bar{b}}=-\sum_{\mathbf{k}'}\vec{V}(\mathbf{k},\mathbf{k}',\mathbf{A})_a^{\;\;\bar{b}}\cdot\left\langle c_{-\mathbf{k}}^{\bar{c}}\vec{\mathbb{\Gamma}}_{\bar{c}}^{\;d}c_{\mathbf{k}'d}\right\rangle.
\end{align}

Now not all of the generators enter in the self-consistent equation above. To understand this, recall that in the single band case, only the x and y components of the pseudospin appear in the right hand side of the self-consistency equation \cite{Anderson1958, Tsuji2015}:
\begin{align}
\label{onebandsce}
\Delta=U\sum_\mathbf{k}(\sigma^x_\mathbf{k}+i\sigma^y_\mathbf{k}).
\end{align}

We follow this clue to determine the relevant generators that enter into the right hand side of the self-consistency equation in the multiband case. To do this we first explicitly separate the particle-hole degrees of freedom from the band degrees of freedom. The particle-hole degrees of freedom can be described by the set of two by two matrices
\begin{align}
\sigma^\alpha=\{\sigma^0,\vec{\sigma}\},
\end{align}
where $\sigma^0$ is the $2\times 2$ identity matrix while the components of $\vec{\sigma}$ are the three Pauli matrices. 

Similarly, the band degrees of freedom are described by the $N\times N$ matrices
\begin{align}
\Theta^\mu=\{\mathbb{\Theta}^0,\vec{\mathbb{\Theta}}\},
\end{align}
where $\mathbb{\Theta}^0$ is an $N\times N$ identity matrix and the $N^2-1$ components of $\vec{\mathbb{\Theta}}$ are the generators of the SU(N) group.

The generators of the full SU(2N) group can then be assembled via
\begin{align}
\label{specialgenerators}
\mathbb{\Lambda}^{\alpha\mu}\equiv\sigma^\alpha\otimes\mathbb{\Theta}^\mu,
\end{align}
for $\alpha=0,\cdot\cdot\cdot,3$ and $\mu=0,\cdot\cdot\cdot,N^2-1$. One of these, specifically $\mathbb{\Lambda}^{00}$, is of course just the $2N\times 2N$ identity matrix.

With the particle-hole degrees of freedom explicitly \lq\lq factored out\rq\rq\;      in the generators eq. \eqref{specialgenerators}, we can now follow eq. \eqref{onebandsce} and write the self-consistency equation for the multiband case as
\begin{align}
\label{sce2}
\Delta_\mathbf{k}(\mathbf{A})_a^{\;\;\bar{b}}=-\sum_{\mathbf{k}'}\vec{V}(\mathbf{k},\mathbf{k}',\mathbf{A})_a^{\;\;\bar{b}}\cdot\left\langle\psi_{\mathbf{k}}^\dagger(\sigma^x-i\sigma^y)\otimes\vec{\mathbb{\Theta}}\psi_{\mathbf{k}}\right\rangle.
\end{align}
That is, the SU(2N) generators $\mathbb{\Lambda}^{1\mu}$ and $\mathbb{\Lambda}^{2\mu}$, for $\mu=0,\cdot\cdot\cdot,N^2-1$ are the generators that are involved in the self-consistency equation.

While the choice eq. \eqref{specialgenerators} makes it clear how to write the self-consistency equation, most SU(2N) generators that are listed on the literature are not of this form. We must then be able to relate $\{(\sigma^x-i\sigma^y)\otimes\vec{\mathbb{\Theta}}\}$ with an arbitrary choice of SU(2N) generators $\{\mathbb{\Gamma}^i|i=0,1,2,\cdot\cdot\cdot 4N^2-1\}$, where for brevity we include the identity matrix in the set. To do this, we expand
\begin{align}
(\sigma^x-i\sigma^y)\otimes\mathbb{\Theta}^\mu=\alpha^\mu_i\mathbb{\Gamma}^i
\end{align}
where
\begin{align}
\alpha^\mu_i=\Tr\left\{\mathbb{\Gamma}_i[(\sigma^x-i\sigma^y)\otimes\mathbb{\Theta}^\mu]\right\}.
\end{align}

We can now write eq. \eqref{sce2} as
\begin{align}
\label{scepseudospin}
\Delta_\mathbf{k}(\mathbf{A})_a^{\;\;\bar{b}}=&-\sum_{\mathbf{k}'}V_\mu(\mathbf{k},\mathbf{k}',\mathbf{A})_a^{\;\;\bar{b}}\alpha_i^\mu\Lambda^i,
\end{align}
where $\Lambda^i=\langle\psi_{\mathbf{k}}^\dagger\mathbb{\Gamma}^i\psi_{\mathbf{k}}\rangle$ are the pseudospin components in terms of the generators $\{\mathbb{\Gamma}^i|i=0,1,2,\cdot\cdot\cdot 4N^2-1\}$.

Equation \eqref{scepseudospin} is our main result of this section. It expresses the self-consistency equation in terms of the multiband pseudospin in terms of arbitrary choice of generators. This and the equations of motion for the pseudospin, form a set of equations that needed to be solved. Lastly, anticipating that this system of equations are to be solved perturbatively, we expand the self-consistency equation in powers of the external field. The zeroth, first, and second orders are, respectively,
\begin{align}
\label{selfconsistent0}
\Delta^{(0)}_\mathbf{k}(\mathbf{A})_a^{\;\;\bar{b}}=&-\sum_{\mathbf{k}'}V^{(0)}_j(\mathbf{k},\mathbf{k}',\mathbf{A})_a^{\;\;\bar{b}}\alpha_i^j\Lambda^{(0)i},\\
\label{selfconsistent1}
\Delta^{(1)}_\mathbf{k}(\mathbf{A})_a^{\;\;\bar{b}}=&-\sum_{\mathbf{k}'}V^{(0)}_j(\mathbf{k},\mathbf{k}',\mathbf{A})_a^{\;\;\bar{b}}\alpha_i^j\Lambda^{(1)i}\nonumber\\
&-\sum_{\mathbf{k}'}V^{(1)}_j(\mathbf{k},\mathbf{k}',\mathbf{A})_a^{\;\;\bar{b}}\alpha_i^j\Lambda^{(0)i},\\
\label{selfconsistent2}
\Delta^{(2)}_\mathbf{k}(\mathbf{A})_a^{\;\;\bar{b}}=&-\sum_{\mathbf{k}'}V^{(0)}_j(\mathbf{k},\mathbf{k}',\mathbf{A})_a^{\;\;\bar{b}}\alpha_i^j\Lambda^{(2)i}\nonumber\\
&-\sum_{\mathbf{k}'}V^{(1)}_j(\mathbf{k},\mathbf{k}',\mathbf{A})_a^{\;\;\bar{b}}\alpha_i^j\Lambda^{(1)i}\nonumber\\
&-\sum_{\mathbf{k}'}V^{(2)}_j(\mathbf{k},\mathbf{k}',\mathbf{A})_a^{\;\;\bar{b}}\alpha_i^j\Lambda^{(0)i},
\end{align}
where the different orders of the interaction matrix and pseudospins are given in the previous sections.

\section{Example: Geometry-induced Leggett mode}
\label{example}
In this section we provide a sample calculation of the formalism that we outlined above. We leave the exploration of more complex case, such as the twisted bilayer graphene superconductor for future work. Here we only focus on the simple case and explore the first order geometric contribution by considering the specific case of the two-band Bernevig-Hughes-Zhang model described by the Hamiltonian \cite{Bernevig2013}
\begin{align}
    H=\sin k_x\sigma_x +\sin k_y \sigma_y +B(2+M-\cos k_x-\cos k_y)\sigma_z.
\end{align}
At the single particle level, this model has a topological phase transition, but is sufficiently simple, which makes it ideal to use in investigating band geometric effects in superconductivity.

We can write the zeroth order pairing potential in the band space and the Berry connection as
\begin{align}
\label{pairingpotential}
    \Delta_\mathbf{k}=
\begin{pmatrix}
    \Delta_{11} & e^{i\alpha}\Delta_{12}\\
    -e^{i\alpha}\Delta_{12} & e^{i\beta}\Delta_{22}
\end{pmatrix}
\end{align}
and
\begin{align}
\mathcal{A}_{\mathbf{k}i}=
\begin{pmatrix}
    a_{\mathbf{k}i} & b_{\mathbf{k}i}+ic_{\mathbf{k}i}\\
    b_{\mathbf{k}i}-ic_{\mathbf{k}i} & d_{\mathbf{k}i}
\end{pmatrix},
\end{align}
where $\Delta_{11}$, $\Delta_{12}$, $\Delta_{22}$, $a_{\mathbf{k}i}$, $b_{\mathbf{k}i}$, $c_{\mathbf{k}i}$, and $d_{\mathbf{k}i}$ are real numbers. Note that the components of the Berry connection are dependent on the parameters $M$ and $B$, but for notational brevity we omit such functional dependence and simply write $a_{\mathbf{k}i}$, $b_{\mathbf{k}i}$ etc. The explicit expressions for the components of the Berry connection is given in the Supplementary Material. We further note that our starting point is a mean-field pairing potential eq. \eqref{pairingpotential} for some suitable interaction matrix that produces such intraband and interband pairing. Since our aim here is to show that there are geometric interband paring fluctuations and leggett mode at the first order, we do not need to specify the microscopic source of the pairing or the interaction matrix. We simply assumed that there is a pairing potential at zeroth order. Such a pairing potential can be intrinsic to the material studied or can come from some bulk superconductor via proximity effect. The angles $\alpha$ and $\beta$ are the phase differences between different band and interband pairings and are therefore physical quantities. Note that we have omitted the superscript zero of $\Delta$ since we will not be concern with higher order corrections.

We can write the different elements of the second term of \eqref{deltai}:
\begin{align}
i\{\Delta_\mathbf{k},\mathcal{A}_\mathbf{k}\}_{11}=&2\Delta_{12}\cos\alpha c_{\mathbf{k}i}\nonumber\\
&+2i(\Delta_{11}a_{\mathbf{k}i}+\Delta_{12}\sin\alpha c_{\mathbf{k}i})\\
i\{\Delta_\mathbf{k},\mathcal{A}_\mathbf{k}\}_{12}=&-\Delta_{11}c_{\mathbf{k}i}-\Delta_{12}(d_{\mathbf{k}i}-a_{\mathbf{k}i})\sin\alpha\nonumber\\
&-\Delta_{22}(b_{\mathbf{k}i}\sin\beta+c_{\mathbf{k}i}\cos\beta)\nonumber\\
&+i[\Delta_{11}b_{\mathbf{k}i}+\Delta_{12}(d_{\mathbf{k}i}+a_{\mathbf{k}i})\cos\alpha\nonumber\\
&+\Delta_{22}(b_{\mathbf{k}i}\cos\beta-c_{\mathbf{k}i}\sin\beta)]\\
i\{\Delta_\mathbf{k},\mathcal{A}_\mathbf{k}\}_{22}=&2\Delta_{12}c_{\mathbf{k}i}\cos\alpha-2\Delta_{22}d_{\mathbf{k}i}\sin\beta\nonumber\\
&+2i(\Delta_{12}c_{\mathbf{k}i}\sin\alpha+\Delta_{22}d_{\mathbf{k}i}\cos\beta).
\end{align}

\begin{figure}[tb]
\centering
\includegraphics[trim={0 0cm 0 0cm},width=1\linewidth]{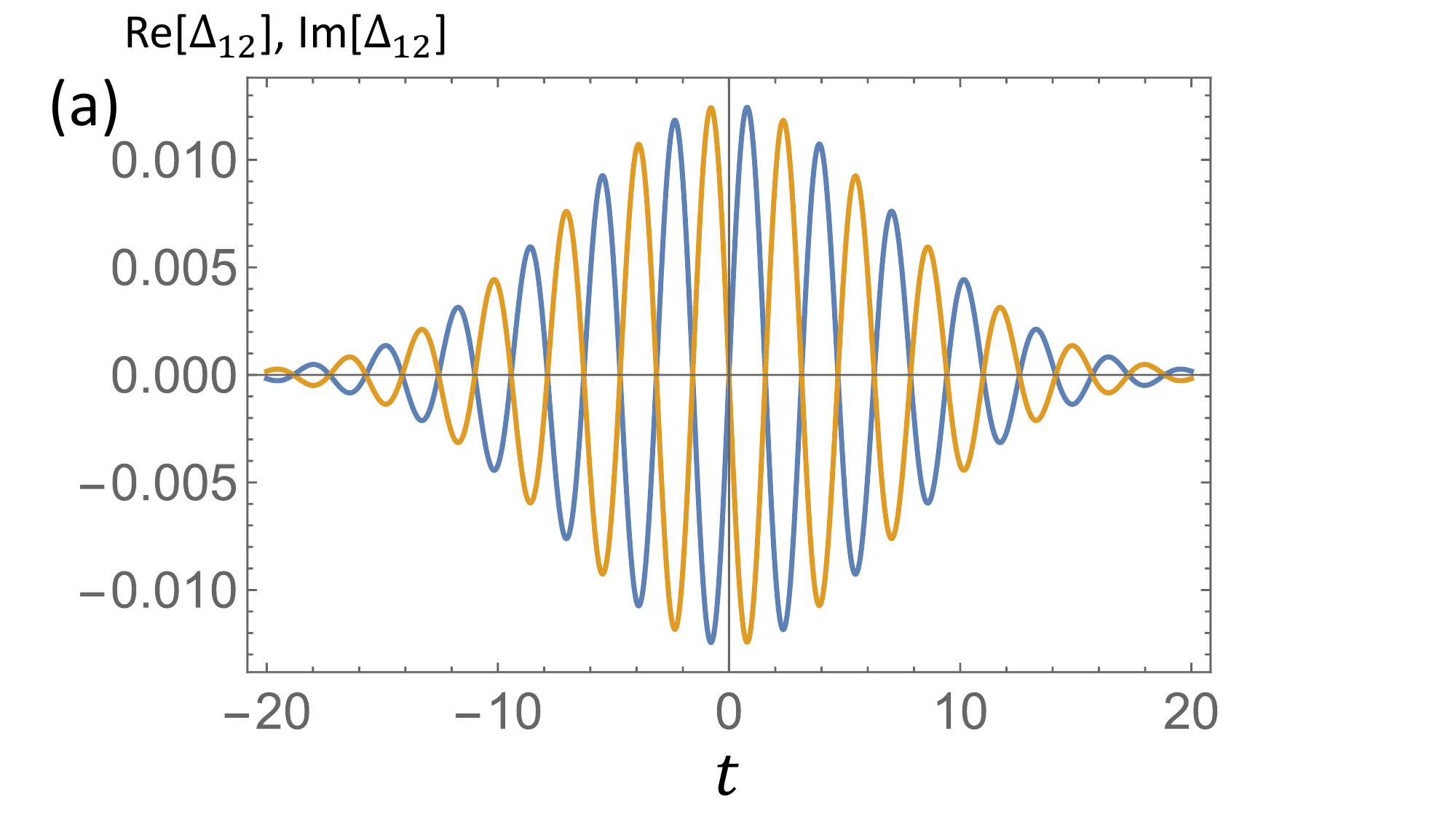}
\includegraphics[trim={0 0cm 0 0.5cm},width=1\linewidth]{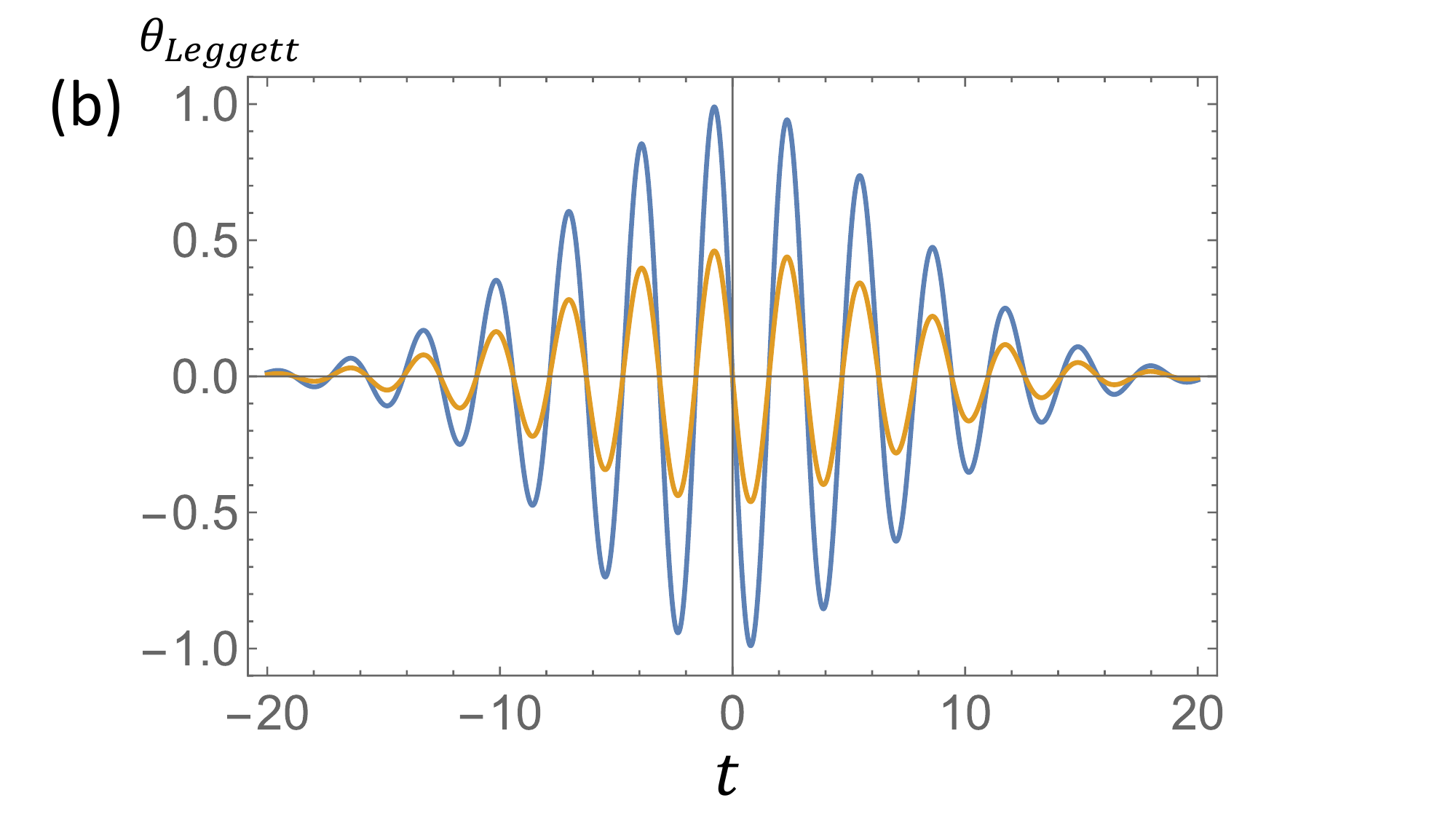}
\caption{(a) Real (blue) and imaginary (orange) parts of the induced interband pairing fluctuations. (b) Leggett modes $\theta_{Leggett}=\delta\theta_{11}-\delta\theta_{12}$ for M=0 (orange) and M=-2 (blue).}
\label{fig:plot1}
\end{figure}

To explore the effects of geometry further, we consider the simpler case where the top band is empty and the pairing only occurs in the bottom band so that to zeroth order $\Delta_{11}\neq 0$ while $\Delta_{12}=\Delta_{22}=0$. The equations above then reduces to
\begin{align}
\label{modes}
i\{\Delta_\mathbf{k},\mathcal{A}_{\mathbf{k},i}\}_{11}=&2i\Delta_{11}a_{\mathbf{k}i}\\
i\{\Delta_\mathbf{k},\mathcal{A}_{\mathbf{k},i}\}_{12}=&-\Delta_{11}c_{\mathbf{k}i}+i\Delta_{11}b_{\mathbf{k}i}\\
i\{\Delta_\mathbf{k},\mathcal{A}_\mathbf{k}\}_{22}=&0.
\end{align}

We see that the induced fluctuation in $\Delta_{11}$ pairing is purely imaginary. Since the zeroth order $\Delta_{11}$ is chosen to be real, an imaginary fluctuation is a pure phase mode. This comes from the expansion
\begin{align}
(\Delta+\delta\Delta)e^{i\delta\theta}\approx \Delta+\delta\Delta+i\Delta\delta\theta ,
\end{align}
showing that the imaginary part comes from the phase fluctuation. This will couple to the external field via Anderson-Higgs mechanism, but this is not our main interest. 

The second equation in \eqref{modes} describes the interband pairing fluctuations induced by the band geometry. Note that in zeroth order we started with $\Delta_{12}=0$. That is, there is no zeroth order interband pairing. In figure \ref{fig:plot1} (a) we show the real and imaginary components of the interband fluctuations using the external pulse field
\begin{align}
\mathbf{A}(t)=\mathbf{A}_0\exp (-t^2/t_0^2)\sin (\omega t),
\end{align}
with $\mathbf{A}_0=\langle 0.1,0.1\rangle$, $t_0=10$, and $\omega=2$. The parameters are $M=-2$, $B=1$, and $\mathbf{k}=\langle 0.1, 0.0\rangle$. At this point in the Brillouin zone the band gap is $\approx 4$ eV, which is much larger than typical pairing $\Delta\leq 10^{-3}$ eV. However, as can be seen from the figure, there is a clear interband mode even if the band gap is large. This can also be seen from the analytic expression, the second equation of \eqref{modes}, that shows no dependence on the band gap. This contribution is purely geometric and is independent of the energy scales of the system. While the phase fluctuation of $\Delta_{11}$ is not a physical quantity, the  fluctuation of the relative phase between $\Delta_{11}$ and $\Delta_{12}$, $\delta\theta_{11}-\delta\theta_{12}$, is a physical quantity, which is the Leggett mode \cite{Leggett1966}. We show this mode in figure \ref{fig:plot1} (b) for different values of the parameter $M$ and $\mathbf{k}=\langle 0.1, 0.0\rangle$. There is a small band gap at this $\mathbf{k}$ value for $M=0$, where the band is gapless at $\mathbf{k}=0$. In contrasts, the band gap at $\mathbf{k}=\langle 0.1, 0.0\rangle$ for $M=-2$ is $\approx 4$ eV. Consistent with the geometric nature of this Leggett mode, its amplitude does not vanish even with the large band gap ($M=-2$, blue curve). In addition, unlike the conventional Leggett mode \cite{Leggett1966}, the geometric Leggett mode that we found here is not proportional to the Fermi velocity. We therefore expect this geometric mode to persist even in the flat band superconductors.

\section{Summary}
\label{summary}
Let us now summarize our results. We start with the generalized Anderson pseudospin theory and the equations that are needed to be solved in using this formalism. First, we have the formal solutions for the pseudospin components eqs. \eqref{lamda0}, \eqref{pseudospinsolution1}, and \eqref{pseudospinsolution2}:
\begin{align}
\vec{\Lambda}_\mathbf{k}^{(0)}=&-\hat{B}^{(0)}_\mathbf{k}(\mathbf{A})\\
\Lambda_\mathbf{k}^{(1)j}=&-2(M_0^{-1})^j_{\;l}f^l_{\;mn}B^{(1)}_\mathbf{k}(\mathbf{A})^{m}\Lambda_\mathbf{k}^{(0)n}\\
\Lambda_\mathbf{k}^{(2)j}=&-2(M_1^{-1})^j_{\;l}f^l_{\;mn}\left(B^{(1)}_\mathbf{k}(\mathbf{A})^{m}\Lambda_\mathbf{k}^{(1)n}\right.\nonumber\\
&\left.+B^{(2)}_\mathbf{k}(\mathbf{A})^{m}\Lambda_\mathbf{k}^{(0)n}\right).
\end{align}
Here we derived the results up to second order, but one can, in principle, obtain the higher orders.

These formal solutions, however, are not yet the final solutions as the right hand side of the equations above contain the unknown order parameter $\Delta^{(j)}$ through the pseudomagnetic field $B_{\mathbf{k}}^{(j)}$. This in turn can be calculated from eq. \eqref{pseudomagneticfield} and then expanded in powers of $\mathbf{A}$ as in eq. \eqref{pseudomagneticfieldexpand}. The different orders of the pseudomagnetic field can be calculated via eqs. \eqref{bcomponent0}, \eqref{bcomponent1}, and \eqref{bcomponent2}.

The equations of motion for the pseudospin must be solved self-consistently. This can be done by substituting the formal solution for the pseudospin components above into the self consistency equation \eqref{scepseudospin}:
\begin{align}
\Delta_\mathbf{k}(\mathbf{A})_a^{\;\;\bar{b}}=&-\sum_{\mathbf{k}'}V_\mu(\mathbf{k},\mathbf{k}',\mathbf{A})_a^{\;\;\bar{b}}\alpha_i^\mu\Lambda^i.
\end{align}

We can also do this order by order by using eqs. \eqref{selfconsistent1} and \eqref{selfconsistent2}, with the expansion of the interaction matrix derived in Section \ref{interaction}. Upon doing this, one ends up with a system of equations with $\Delta^{(n)}_{\mathbf{k},{dir}}$ as the unknowns, where $n$ labels the order. Equations \eqref{deltai} and \eqref{deltaij} then gives the first and second order fluctuations to the pairing potential. These fluctuations contain contributions that are purely geometric in nature. In the special case where one projects to a single band case, these fluctuations were shown to survive even in the flat band limit \cite{Villegas2021}. At first order, we also saw that the geometric contribution induces interband pairing fluctuations that are independent of the band gap. Since these pairing fluctuations are written in the band basis, the off-diagonal solutions give the interband pairing fluctuations, while the diagonal terms give the intraband fluctuations. When the zeroth order pairing matrix in the band basis is known, the fluctuations in magnitude or radial fluctuations will give us the Higgs mode, while the fluctuations of the phases for different band pairings will give us the Leggett modes. 

We provided a simple first order calculation using the Bernevig-Hughes-Zhang model. We have shown that the interband pairing fluctuation and the Leggett mode persists even when the band gap is larger than the energy scale of the pairing.

It would be interesting to explore the consequences of our theory to specific models of multiband superconductors, such as the layered systems. A full calculation would require solving the self-consistency equation and integrating either over the Fermi surface, or the entire band in the case of flat bands. One may also study the interplay of dispersive and flat bands in the interband pairing and the Leggett modes by using other toy models that are known to have nontrivial band geometry \cite{Hofmann2020, Mitscherling2022}. We leave all such detailed explorations for future work.

\begin{acknowledgments}
The author would like to thank B. Yang for the insightful discussions. This work was funded by the UP System Balik PHD Program (OVPAA-BPhD-2022-06).
\end{acknowledgments}
\bibliography{Arxiv}
\bibliographystyle{apsrev4-2}

\begin{widetext}

\appendix

\newpage

\section{SUPPLEMENTAL MATERIAL: Berry connection of the Bernevig-Hughes-Zhang model.}

\maketitle

In this Supplemental Material, we provide the explicit expressions for the Bloch eigenfunctions and the Berry connection components of the Bernevig-Hughes-Zhang model that we used in the main text. These give explicit dependence of these quantities on the parameters $B$ and $M$.

The normalized eigenfunctions are 
\begin{align}
\psi_1(B,M,k_x,k_y)=&\left(\frac{-\sin k_x+i \sin k_y}{\sqrt{2} f^+(B,M,k_x,k_y)},\frac{f^+(B,M,k_x,k_y)}{\sqrt{2} f(B,M,k_x,k_y)}\right)^T\\
\psi_2(B,M,k_x,k_y)=&\left(\frac{\sin k_x-i \sin k_y}{\sqrt{2} f^-(B,M,k_x,k_y)},\frac{f^-(B,M,k_x,k_y)}{\sqrt{2} f(B,M,k_x,k_y)}\right)^T.
\end{align}

Here
\begin{align}
f(B,M,k_x,k_y)\equiv &\sqrt{(M-B (\cos k_x+\cos k_y))^2+\sin ^2k_x+\sin ^2k_y}\\
f^\pm(B,M,k_x,k_y)\equiv&\sqrt{f(B,M,k_x,k_y) \left[f(B,M,k_x,k_y)\mp B (\cos k_x+\cos k_y)\pm M\right]}
\end{align}

From this we can calculate the components of the Berry connection
\begin{align}
\vec{\mathcal{A}}_{ij}=i\psi_i(B,M,k_x,k_y)^\dagger \nabla_\mathbf{k}\psi_j(B,M,k_x,k_y).
\end{align}

The results are
\begin{align}
[\mathcal{A}_x]_{11}=&
\frac{i}{2f^+(B,M,k_x,k_y)^3}\bigg[-\frac{\partial f(B,M,k_x,k_y)}{\partial k_x}\frac{ f^+(B,M,k_x,k_y)^5}{f(B,M,k_x,k_y)^3}+\frac{\partial f^+(B,M,k_x,k_y)}{\partial k_x}\frac{f^+(B,M,k_x,k_y)^4}{f(B,M,k_x,k_y)^2}\nonumber\\
&-\left(\sin ^2k_x+\sin ^2k_y\right) \frac{\partial f^+(B,M,k_x,k_y)}{\partial k_x}+\cos (k_x) (\sin k_x+i \sin k_y) f^+(B,M,k_x,k_y)\bigg]\\
[\mathcal{A}_x]_{12}=&
i \bigg[-\frac{\partial f(B,M,k_x,k_y)}{\partial k_x}\frac{ f^-(B,M,k_x,k_y) f^+(B,M,k_x,k_y)}{2 f(B,M,k_x,k_y)^3}+\frac{f^-(B,M,k_x,k_y)}{\partial k_x}\frac{ f^+(B,M,k_x,k_y)}{2 f(B,M,k_x,k_y)^2}\nonumber\\
&+\frac{\left(\sin ^2k_x+\sin ^2k_y\right) \partial_{k_x}f^-(B,M,k_x,k_y)-(\cos k_x) (\sin k_x+i \sin k_y) f^-(B,M,k_x,k_y)}{2f^-(B,M,k_x,k_y)^2f^+(B,M,k_x,k_y)}\bigg]\\
[\mathcal{A}_x]_{21}=&i \bigg[\frac{f^-(B,M,k_x,k_y)}{2 f(B,M,k_x,k_y)^3} \left(f(B,M,k_x,k_y) \frac{\partial f^+(B,M,k_x,k_y)}{\partial k_x}-\frac{\partial f(B,M,k_x,k_y)}{\partial k_x} f^+(B,M,k_x,k_y)\right)\nonumber\\
&+\frac{\left(\sin ^2k_x+\sin ^2k_y\right) \partial_{k_x}f^+(B,M,k_x,k_y)-\cos k_x (\sin k_x+i \sin k_y)f^+(B,M,k_x,k_y)}{2f^-(B,M,k_x,k_y)f^+(B,M,k_x,k_y)^2}\bigg]\\
[\mathcal{A}_x]_{22}=&\frac{i }{2 f^-(B,M,k_x,k_y)^3}\bigg[-\frac{\partial f(B,M,k_x,k_y)}{\partial k_x}\frac{ f^-(B,M,k_x,k_y)^5}{f(B,M,k_x,k_y)^3}+\frac{\partial f^-(B,M,k_x,k_y)}{\partial k_x}\frac{ f^-(B,M,\text{kx},k_y)^4}{f(B,M,k_x,k_y)^2}\nonumber\\
&-\left(\sin ^2k_x+\sin ^2k_y\right) \frac{\partial f^-(B,M,k_x,k_y)}{\partial k_x}+\cos k_x (\sin k_x+i \sin k_y) f^-(B,M,k_x,k_y)\bigg]\\
[\mathcal{A}_y]_{11}=&\frac{i}{2 f^+(B,M,k_x,k_y)^3}\bigg[-\frac{\partial f(B,M,k_x,k_y)}{\partial k_y}\frac{ f^+(B,M,k_x,k_y)^5}{f(B,M,k_x,k_y)^3}+\frac{\partial f^+(B,M,k_x,k_y)}{\partial k_y}\frac{ f^+(B,M,k_x,k_y)^4}{f(B,M,k_x,k_y)^2}\nonumber\\
&-\left(\sin ^2k_x+\sin ^2k_y\right) \frac{\partial f^+(B,M,k_x,k_y)}{\partial k_y}+\cos k_y (\sin k_y-i \sin k_x) f^+(B,M,k_x,k_y)\bigg]\\
[\mathcal{A}_y]_{12}=&
i \bigg[-\frac{\partial f(B,M,k_x,k_y)}{\partial k_y}\frac{f^-(B,M,k_x,k_y) f^+(B,M,k_x,k_y)}{2 f(B,M,k_x,k_y)^3}+\frac{f^-(B,M,k_x,k_y)}{\partial k_y}\frac{f^+(B,M,k_x,k_y)}{2 f(B,M,k_x,k_y)^2}\nonumber\\
&+\frac{1}{2 f^-(B,M,k_x,k_y)^2 f^+(B,M,k_x,k_y)}\bigg(\left(\sin ^2k_x+\sin ^2k_y\right) \frac{\partial f^-(B,M,k_x,k_y)}{\partial k_y}\nonumber\\
&-\cos k_y (\sin k_y-i \sin k_x) f^-(B,M,k_x,k_y)\bigg)\bigg]\\
[\mathcal{A}_y]_{21}=&i \frac{f^-(B,M,k_x,k_y)}{2 f(B,M,k_x,k_y)^3}\bigg[f(B,M,k_x,k_y) \frac{\partial f^+(B,M,k_x,k_y)}{\partial k_y}-\frac{\partial f(B,M,k_x,k_y)}{\partial k_y} f^+(B,M,k_x,k_y)\bigg]\nonumber\\
&+\frac{i}{2 f^-(B,M,k_x,k_y) f^+(B,M,k_x,k_y)^2}\bigg[\left(\sin ^2k_x+\sin ^2k_y\right) \frac{\partial f^+(B,M,k_x,k_y)}{\partial k_y}\nonumber\\
&-\cos k_y (\sin k_y-i \sin k_x) f^+(B,M,k_x,k_y)\bigg]\\
[\mathcal{A}_y]_{22}=&\frac{i }{2 f^-(B,M,k_x,k_y)^3}\bigg[-\frac{\partial f(B,M,k_x,k_y)}{\partial k_y}\frac{f^-(B,M,k_x,k_y)^5}{f(B,M,k_x,k_y)^3}+\frac{\partial f^-(B,M,k_x,k_y)}{\partial k_y}\frac{f^-(B,M,k_x,k_y)^4}{f(B,M,k_x,k_y)^2}\nonumber\\
&-\left(\sin ^2k_x+\sin ^2k_y\right) \frac{\partial f^-(B,M,k_x,k_y)}{\partial k_y}+\cos k_y (\sin k_y-i \sin k_x) f^-(B,M,k_x,k_y)\bigg].
\end{align}

\end{widetext}

\end{document}